\documentclass[10pt,pre,aps,groupaddress]{revtex4-1}
\usepackage{epsfig,amsmath,graphicx,amssymb,overpic}
\usepackage{color}
\def\be{\begin{equation}}
\def\ee{\end{equation}}
\def\bee{\begin{eqnarray}}
\def\ene{\end{eqnarray}}
\def\bes{\begin{subequations}}
\def\ees{\end{subequations}}

\begin{document}

\title{Financial Rogue Waves Appearing in the Coupled Nonlinear \\
\vspace{0.1in} Volatility and Option Pricing Model}
\author{\vspace{0.2in} {\bf Zhenya Yan}\footnote{Email address: zyyan@mmrc.iss.ac.cn; zyyan\_math@yahoo.com.} }

\affiliation{\vspace{0.1in}  Key Laboratory of Mathematics Mechanization,
 Institute of Systems Science, AMSS, Chinese Academy of Sciences,
Beijing 100190, China }

\vspace{9in}

\begin{abstract}
\baselineskip=18pt

\vspace{0.2in} {\bf Abstract.}  The coupled nonlinear volatility and option pricing model presented recently by Ivancevic is investigated, which generates a leverage effect, i.e., stock volatility is (negatively)
correlated to stock returns, and can be regarded as a coupled nonlinear wave alternative of the Black-Scholes option pricing model. In this short report, we analytically propose the two-component financial rogue waves
of the coupled nonlinear volatility and option pricing model without an embedded $w-$learning. Moreover, we exhibit their dynamical behaviors for chosen different parameters. The two-component financial rogue wave solutions
may be used to describe the possible physical mechanisms for the rogue wave phenomena and
to further excite the possibility of relative researches and potential applications of rogue waves
in the financial markets and other related fields. \\


\end{abstract}


\maketitle


\baselineskip=15pt

\section{Introduction}

The use of the term freak wave was first introduced to the scientific community by Draper~\cite{rwd,rgd2}.
Freak waves are also called rogue waves, giant waves, monster waves, killer waves, extreme waves, abnormal waves,
huge waves, super waves, or gigantic waves, which are formed due to the interaction of the nonlinearity and dispersion in the wave propagation.
Nowadays, there is no unique definite for the rogue waves, but, a wave called a rogue wave means that the
wave height $H$ (i.e., distance from trough to crest) or the crest height $\eta_c$ (i.e., distance from
mean sea level to crest) exceeds a certain threshold related to the sea state~\cite{rgd2}. Generally speaking,
the common criteria of rogue waves is that $H>2H_s$ or $\eta_c>1.5H_s$, where $H_s$ stands for the significant wave height~\cite{rgd2}.

Rogue waves, as a special type of solitary waves, have drawn much attention
in some fields of nonlinear science such as oceanics~\cite{rgd2,RW1,RW2,RW3,RW4,RW5}, nonlinear optics~\cite{ORW, exp1, exp2, ABC1,ABC2,ABC3,BKA,ypla10,ypla102}, Bose-Einstein condensations~\cite{BRW, ypre10}, atmospherics~\cite{arg}, and even finance~\cite{frg}. The first-order rational
solution of the self-focusing nonlinear Schr\"odinger (NLS) equation was first found by Peregrine~\cite{PS}
to describe the rogue wave phenomenon, which was known as Peregrine soliton (or Peregrine
breather) and could be regarded as a limiting case of the Ma breather when the breathing period approaches to infinity~\cite{Ma,PST} or of the Akhmediev breather as the spatial period tends to infinity~\cite{AK,PST}.

More recently,
multi-rogue wave solutions of the self-focusing NLS equation were also found
on basis of the deformed Darboux transformation in~\cite{ABC1, ABC2}. Moreover,
nonautonomous  rogue wave solutions have also been found for the generalized NLS equations with variable coefficients in one-dimensional space~\cite{ypla10,ypla102} and in three-dimensional spaces~\cite{ypre10} based on the similarity analysis idea. The term `rogue (freak) waves' was called `rogons (freakons)' if they reappeared virtually unaffected in size or shape shortly after their interactions~\cite{ypla10}.
In nonlinear optics, optical rogue waves have been observed in two different experiments, based on two nonlinear optical models, i.e., the higher-order NLS equation~\cite{exp1} and the NLS equation~\cite{exp2}, respectively. Moreover, they have verified that optical rogue waves reveal an approach to exert
maximal control over a nonlinear system with minimal effort and can be used to produce highly stable
supercontinuum~\cite{exp1}.

As is well known, the celebrated Black-Scholes (alias as Black-Scholes-Merton) option pricing model~\cite{BS, BS2}
 \bee
  \frac{\partial C(S,t)}{\partial t}+\frac12\sigma^2 S^2\frac{\partial^2 C(S,t)}{\partial S^2}
   +rS\frac{\partial C(S,t)}{\partial S}-rC(S,t)=0, \ene
was presented, based on the the geometric Brownian motion (i.e. the
stochastic differential equation) $dS=\mu Sdt+\sigma dW(t)$
satisfied by the stock (asset) price $S$ and the It$\hat{\rm o}$
lemma~\cite{Ito}, where $C(S,t)$ is the value
of European call option on the asset price $S$ at time $t$, $\mu$
is the instantaneous mean return, $\sigma$  is the stock
volatility, $W$ is a Wiener process,  and $r$ is the risk-free
interest rate. This model has drawn much attention and also opens
the new area of study in financial mathematics or financial engineering.
The Black-Scholes model can be widely used for valuing the pricing of European-style options, but it cannot be used for valuing other exotic types of options such as American Options or Asian Options as it cannot incorporate
exercise features or any path dependencies~\cite{book1,book2,book3}.

More recently, based on the modern adaptive markets
hypothesis due to Lo~\cite{Lo, Lo2}, Elliott wave market
theory~\cite{Elliot, Elliot2}, and quantum neural computation
approach~\cite{Ivan}, Ivancevic presented a novel nonlinear option pricing
model (called the {\it Ivancevic option pricing model})~\cite{Ivan2}
 \bee
  \label{nls}
   i\frac{\partial \psi(S,t)}{\partial t}=-\frac12\sigma\frac{\partial^2 \psi(S,t)}{\partial S^2}
    -\beta\left|\psi(S,t)\right|^2\psi(S,t), \qquad i=\sqrt{-1} \ene
in order to satisfy efficient and behavioral markets, and their
essential nonlinear complexity, where $\psi(S,t)$ denotes the
{\it option-price wave function}, the dispersion frequency
coefficient $\sigma$ is the volatility (which can be either a
constant or stochastic process itself), the Landau coefficient
$\beta=\beta(r,w)$ represents the adaptive market potential. Some
periodic wave solutions of Eq. (\ref{nls}) have been
obtained~\cite{Ivan2}. Recently, we first presented the two financial rogue wave solutions of the Ivancevic option pricing model (\ref{nls}) and illustrated
their dynamics~\cite{frg}, which may explain real financial crisis/storms (e.g., 1997 Asian financial crisis/storm and the current global financial crisis/storm).
In addition, for the aim of including a controlled stochastic volatility into the adaptive
Ivancevic option pricing model (\ref{nls}), the adaptive, symmetrically coupled, volatility and option pricing
model (with interest rate $r$ and Hebbian learning rate $c$) has been presented, which represents a bidirectional
spatio-temporal associative memory~\cite{Ivan2, Ivan3}. In the following, we would like to study the financial rogue waves of the coupled nonlinear model.

 \section{The coupled nonlinear volatility and option pricing model}

By introducing a controlled stochastic volatility into the adaptive-wave model (\ref{nls}), Ivancevic also
presented the coupled nonlinear volatility and option pricing model in the form~\cite{Ivan2,Ivan3}
 \bes \label{nlsc} \bee
    \label{nlsc1} && i\frac{\partial \sigma(S,t)}{\partial t}=-\frac12\frac{\partial^2 \sigma(S,t)}{\partial S^2}
    -\beta(r,w)\Big[\big|\,\sigma(S,t)\big|^2+\big|\,\psi(S,t)\big|^2\Big]\sigma(S,t), \\
\label{nlsc2}  && i\frac{\partial \psi(S,t)}{\partial t}=-\frac12\frac{\partial^2 \psi(S,t)}{\partial S^2}
    -\beta(r,w)\Big[\big|\,\sigma(S,t)\big|^2+\big|\,\psi(S,t)\big|^2\Big]\psi(S,t),\qquad i=\sqrt{-1}
    \ene \ees
to generate a leverage effect, i.e., stock volatility is (negatively)
correlated to stock returns~\cite{nlsc, nlsc2}, where Eqs.~(\ref{nlsc1}) and (\ref{nlsc2}) are called the volatility model and option pricing model, respectively,  $\psi(S,t)$ denotes the {\it option pricing wave function}, playing the role of a nonlinear coefficient in
the volatility model (\ref{nlsc1}), $\sigma(S,t)$ denotes the {\it volatility wave function}, standing for
a nonlinear coefficient in the option pricing model (\ref{nlsc2}). Both processes thus evolve in a common self-organizing market heat potential,
so they effectively represent an adaptively controlled Brownian behavior of a hypothetical financial market,
 $\beta(r,w)=r\sum_{j=1}^Nw_jg_j$ with the adaptation equation $\dot{w}_j=-w_j+c|\sigma|g_j|\psi|$~\cite{Ivan2}. Eqs.~(\ref{nlsc1}) and (\ref{nlsc2}) without an embedded $w-$learning (i.e., for constant
$\beta=r$-the interest rate), possessed the solitary wave solutions~\cite{Ivan2}.

To the best of our knowledge, the analytical financial rogue waves of Eqs.~(\ref{nlsc1}) and (\ref{nlsc2}) were not reported before. In the following, we would like to investigate the two-component
financial rogue wave solutions of Eqs.~(\ref{nlsc1}) and (\ref{nlsc2}) without an embedded $w-$learning,
which may play an important role to explain some occurred financial crisis/storms (e.g., 1997 Asian financial crisis/storm and the current global financial crisis/storm).

 \section{Financial rogue waves (rogons)}

Here, based on the similarity analysis and the approach developed in Refs.~\cite{ABC1, ABC2, ypla10, ypre10, frg}, we
show that the coupled nonlinear volatility and option pricing model (\ref{nlsc1}) and (\ref{nlsc2})
 without an embedded $w-$learning also admits the financial multi-rogon (rogue wave) solutions, which
may be used to describe the possible formation mechanisms for
rogue wave phenomenon in financial markets. In the following, we only give the first
two representative financial rogue wave (rogon) solutions of Eqs.~(\ref{nlsc1}) and (\ref{nlsc2}) without an embedded $w-$learning .

\subsection{Financial one-rogue waves (rogons)}

\begin{figure}[!ht]
\begin{center}
{\scalebox{0.7}[0.7]{\includegraphics{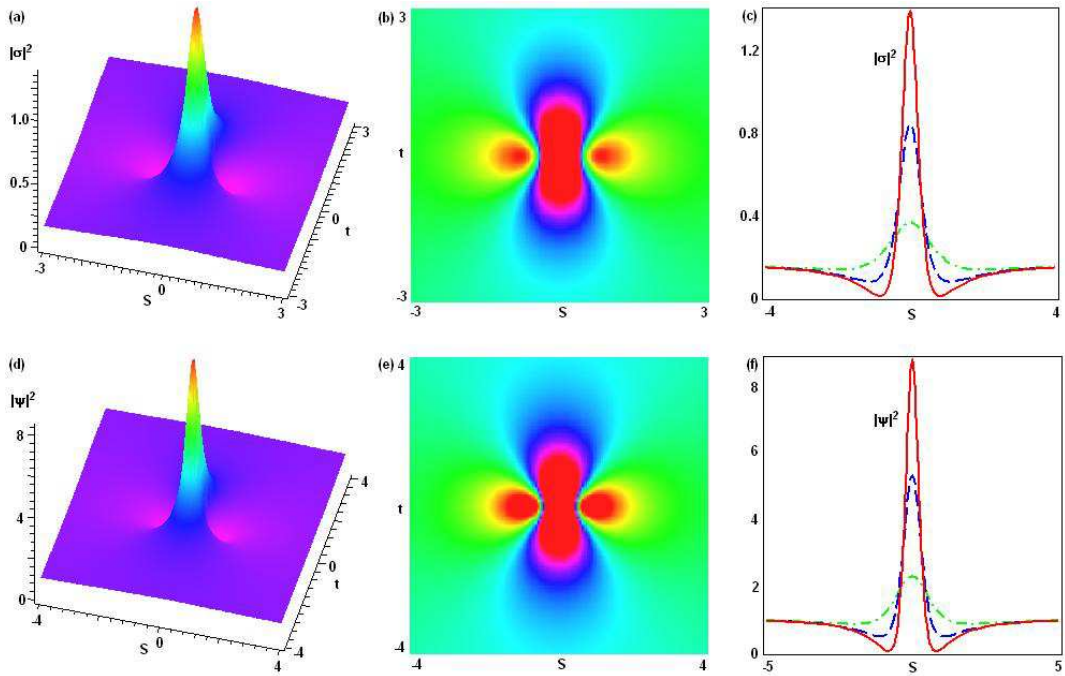}}}
\end{center}
\vspace{-0.2in} \caption{\small (color online). Financial rogue wave
propagations for the coupled volatility wave function $\sigma_1(S,t)$
and option pricing  wave function $\psi_1(S,t)$ of the financial one-rogon solutions given by Eqs.~(\ref{solu1a}) and (\ref{solu1b}) for parameters $\alpha=1.5,\, \beta=1,\, a=2,\, b=5$, and $k=0$. (a) The intensity distribution $|\sigma_1(S,t)|^2$; (b) the density distridution of $|\sigma_1(S,t)|^2$; (c) The intensity distribution $|\sigma_1(S,t)|^2$ for time $t=0$ (solid line), $t=0.4$ (dashed line), $t=1.0$ (dashed-dotted line);  (d) The intensity distribution $|\psi_1(S,t)|^2$; (e) the density distribution of  $|\psi_1(S,t)|^2$; (f) The intensity distribution $|\psi_1(S,t)|^2$ for time $t=0$ (solid line), $t=0.4$ (dashed line), $t=1.0$ (dashed-dotted line). }
\label{fig:1}
\end{figure}

\begin{figure}[!ht]
\begin{center}
\vspace{0.05in} {\scalebox{0.7}[0.75]{\includegraphics{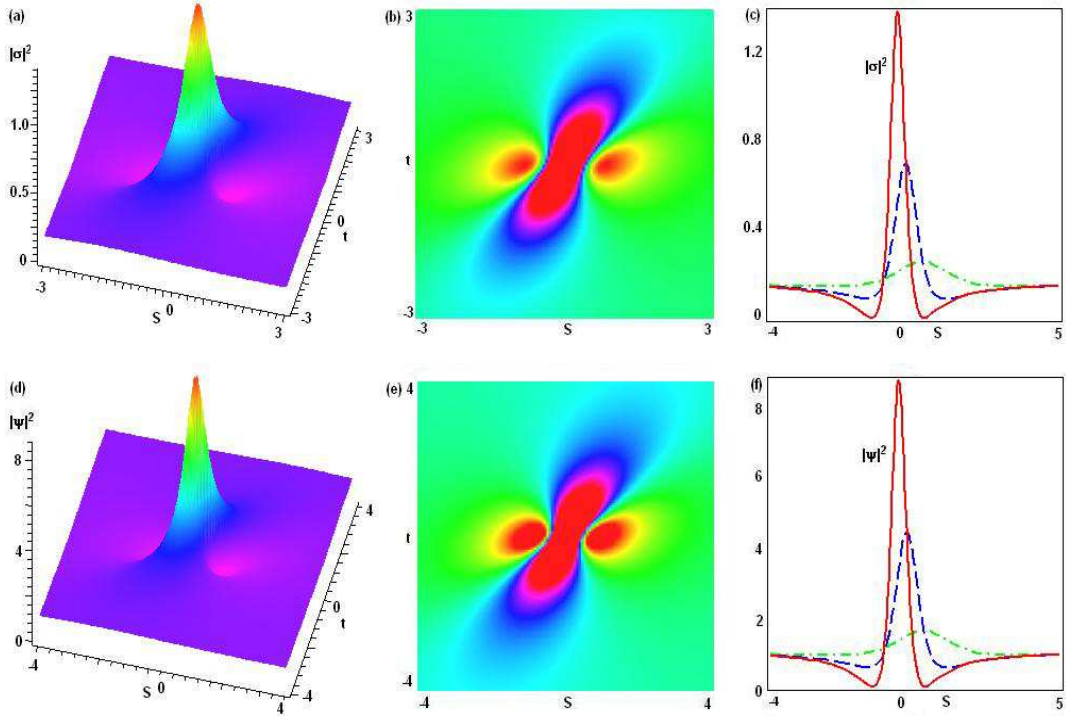}}}
\end{center}
\vspace{-0.2in} \caption{\small (color online). Financial rogue wave
propagations for the coupled volatility wave function $\sigma_1(S,t)$
and option pricing  wave function $\psi_1(S,t)$ of the financial one-rogon solutions given by Eqs.~(\ref{solu1a}) and (\ref{solu1b}) for parameters $\alpha=1.5,\, \beta=1,\, a=2,\, b=5$, and $k=0.5$. (a) The intensity distribution $|\sigma_1(S,t)|^2$; (b) the density distridution of $|\sigma_1(S,t)|^2$; (c) The intensity distribution $|\sigma_1(S,t)|^2$ for time $t=0$ (solid line), $t=0.4$ (dashed line), $t=1.0$ (dashed-dotted line);  (d) The intensity distribution $|\psi_1(S,t)|^2$; (e) the density distribution of  $|\psi_1(S,t)|^2$; (f) The intensity distribution $|\psi_1(S,t)|^2$ for time $t=0$ (solid line), $t=0.4$ (dashed line), $t=1.0$ (dashed-dotted line).}\label{fig:2}
\end{figure}

 The financial one-rogon solution
of Eqs.~(\ref{nlsc1}) and (\ref{nlsc2})  without an embedded $w-$learning  for the coupled volatility wave function $\sigma(S,t)$ and option pricing wave function $\psi(S,t)$ can be written as
 \bes \label{solu1}
 \bee
 \label{solu1a}
 && \sigma_1(S,t)=\frac{\alpha a}{\sqrt{2\beta (a^2+b^2)}}\left[1-\frac{4(1+i\,\alpha^2t)}
  {1+2\alpha^2(S-kt)^2+\alpha^4t^2}\right]
 e^{i[kS+(\alpha^2-k^2)t/2]}, \\
\label{solu1b} && \psi_1(S,t)=\frac{\alpha b}{\sqrt{2\beta (a^2+b^2)}}\left[1-\frac{4(1+i\,\alpha^2t)}
  {1+2\alpha^2(S-kt)^2+\alpha^4t^2}\right]
 e^{i[kS+(\alpha^2-k^2)t/2]},
  \ene \ees
in terms of the complex rational functions of the stock price $S$
and time $t$, which involves five free parameters $\alpha,\, \beta, \, a,\, b,$
and $k$ to manage the different types of financial one-rogue wave
propagations, whose intensity distributions $|\sigma_1(S,t)|^2$ and $|\psi_1(S,t)|^2$ are displayed in
Figs.~\ref{fig:1} and \ref{fig:2} for the chosen adaptive market
potential $\beta=1$, the scaling $\alpha=1.5$, amplitude parameters $a=2,\, b=5$, and the gauge $k=0,\,
0.5$. Notice that time $t$ in these two figures can be chosen to be negative
since the solution is invariant under the translation transformation $t\rightarrow t-t_0$.

\subsection{Financial two-rogue waves (rogons)}

\begin{figure}[!ht]
\begin{center}
\vspace{0.1in}{\scalebox{0.7}[0.72]{\includegraphics{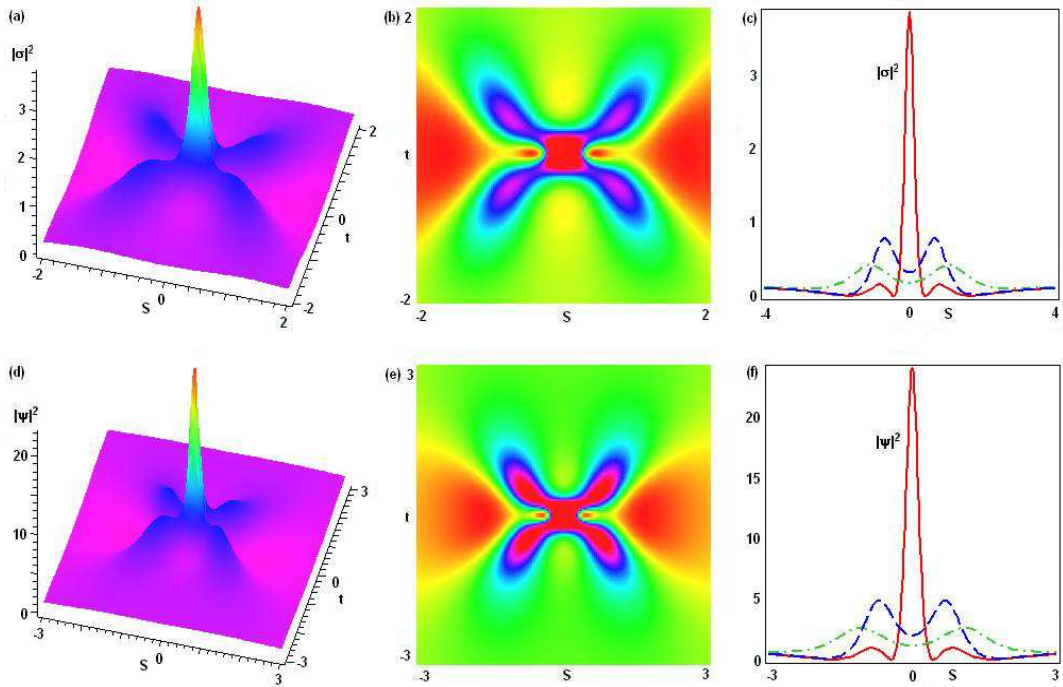}}}
\end{center}
\vspace{-0.1in} \caption{\small (color online). Financial rogue wave
propagations for the coupled volatility wave function $\sigma_2(S,t)$
and option pricing  wave function $\psi_2(S,t)$ of the financial two-rogon solutions given by Eqs.~(\ref{solu2a}) and (\ref{solu2b}) for parameters $\alpha=1.5,\, \beta=1,\, a=2,\, b=5$, and $k=0$. (a) The intensity distribution $|\sigma_2(S,t)|^2$; (b) the density distridution of $|\sigma_2(S,t)|^2$; (c) The intensity distribution $|\sigma_2(S,t)|^2$ for time $t=0$ (solid line), $t=0.4$ (dashed line), $t=1.2$ (dashed-dotted line);  (d) The intensity distribution $|\psi_2(S,t)|^2$; (e) the density distribution of  $|\psi_2(S,t)|^2$; (f) The intensity distribution $|\psi_2(S,t)|^2$ for time $t=0$ (solid line), $t=0.4$ (dashed line), $t=1.2$ (dashed-dotted line). }\label{fig:3}
\end{figure}

\begin{figure}[!ht]
\begin{center}
\vspace{0.1in} {\scalebox{0.7}[0.72]{\includegraphics{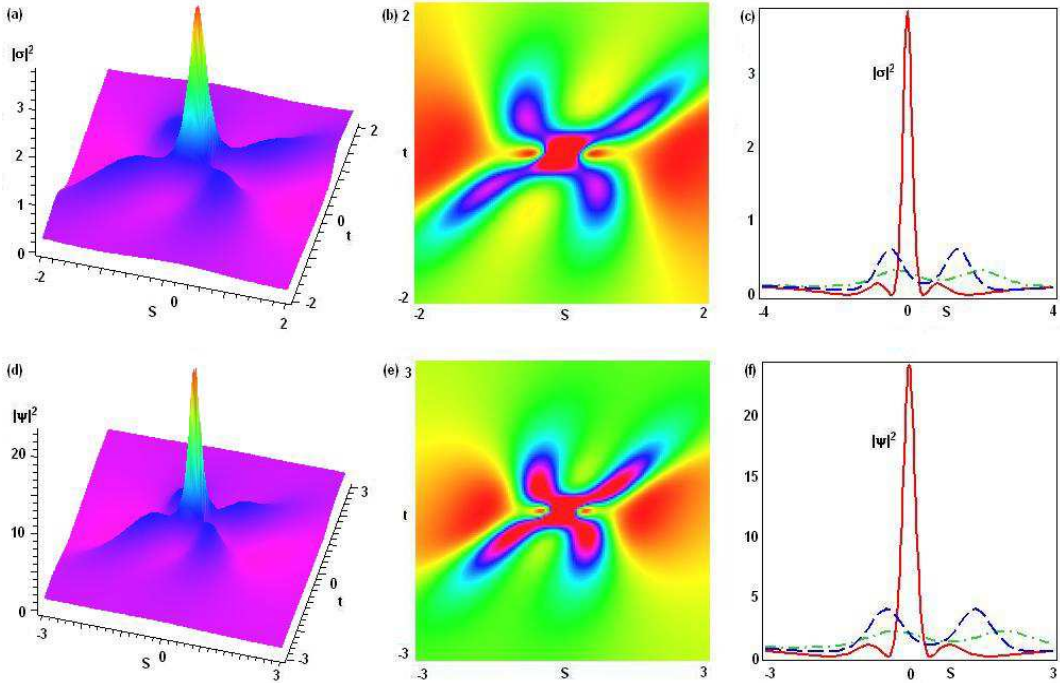}}}
\end{center}
\vspace{-0.1in} \caption{\small (color online). Financial rogue wave
propagations for the coupled volatility wave function $\sigma_2(S,t)$
and option pricing  wave function $\psi_2(S,t)$ of the financial two-rogon solutions given by Eqs.~(\ref{solu2a}) and (\ref{solu2b}) for parameters $\alpha=1.5,\, \beta=1,\, a=2,\, b=5$, and $k=0.5$. (a) The intensity distribution $|\sigma_2(S,t)|^2$; (b) the density distridution of $|\sigma_2(S,t)|^2$; (c) The intensity distribution $|\sigma_2(S,t)|^2$ for time $t=0$ (solid line), $t=0.8$ (dashed line), $t=1.5$ (dashed-dotted line);  (d) The intensity distribution $|\psi_2(S,t)|^2$; (e) the density distribution of  $|\psi_2(S,t)|^2$; (f) The intensity distribution $|\psi_2(S,t)|^2$ for time $t=0$ (solid line), $t=0.8$ (dashed line), $t=1.5$ (dashed-dotted line).  }\label{fig:4}
\end{figure}

\quad Similarly, based on the symmetry analysis, we can obtain the financial two-rogon solutions of Eqs.~(\ref{nlsc1}) and (\ref{nlsc2}) without an embedded $w-$learning for the coupled volatility wave function $\sigma(S,t)$ and option pricing  wave function $\psi(S,t)$ in the form
 \bes \label{solu2}
 \bee
 \label{solu2a}
 && \sigma_2(S,t)=\frac{\alpha a}{\sqrt{2\beta (a^2+b^2)}}
 \left[1+\frac{P_2(S,t)-(1/2)\,i\,\alpha^2 t\,Q_2(S,t)}{H_2(S,t)}\right]
 e^{i[kS+(\alpha^2-k^2)t/2]},  \\
\label{solu2b} && \psi_2(S,t)=\frac{\alpha b}{\sqrt{2\beta (a^2+b^2)}}
 \left[1+\frac{P_2(S,t)-(1/2)\,i\,\alpha^2 t\,Q_2(S,t)}{H_2(S,t)}\right]
 e^{i[kS+(\alpha^2-k^2)t/2]},
  \ene \ees
with these above-mentioned functions $P_2(S,t), \ Q_2(S,t)$ and $H_2(S,t)$ being
of  polynomials in  the stock price $S$ and time $t$ in the form
 \bee
 \label{solu2c}
  \begin{array}{l}
 P_2(S,t)=\displaystyle -\frac{\alpha^4}{2}(S- kt)^4-\frac{3\alpha^6}{2}(S-kt)^2 t^2 -\frac{5\alpha^8}{8}t^4-\frac{3\alpha^2}{2}(S-kt)^2-\frac{9\alpha^4}{4}t^2+\frac38,
\vspace{0.1in}\cr
  Q_2(S,t)=\displaystyle \alpha^4(S- kt)^4+\alpha^6(S-kt)^2t^2
 +\frac{\alpha^8}{4}t^4 -3\alpha^2(S- kt)^2+\frac{\alpha^4}{2}t^2-\frac{15}{4},
\vspace{0.1in}\cr
H_2(S,t)=\displaystyle  \frac{\alpha^6}{12}(S- kt)^6+\frac{\alpha^8}{8}(S-kt)^4t^2
  +\frac{\alpha^{10}}{16}(S- kt)^2t^4+\frac{\alpha^{12}}{96}t^6+\frac{\alpha^4}{8}(S-kt)^4
\vspace{0.08in}\cr
 \qquad\qquad\quad\quad  \displaystyle
 -\frac{3\alpha^6}{8}(S-kt)^2t^2
 +\frac{9\alpha^8}{32}t^4
  +\frac{9\alpha^2}{16}(S- kt)^2+\frac{33\alpha^4}{32}t^2+\frac{3}{32}, \end{array} \ene
which contain five free parameters $\alpha,\, \beta, \, a,\, b,$
and $k$ to manage the different types of financial two-rogue wave
propagations whose intensity distributions $|\sigma_2(S,t)|^2$ and $|\psi_2(S,t)|^2$ are depicted in
Figs.~\ref{fig:3} and \ref{fig:4} for the chosen adaptive market
potential $\beta=1$, the scaling $\alpha=1.5$, amplitude parameters $a=2,\, b=5$ and the gauge $k=0,\,
0.5$.

 \section{Conclusions}

 In conclusion, based on the symmetry analysis, we have investigated  the coupled nonlinear volatility and option pricing model (\ref{nlsc1}) and (\ref{nlsc2}) without an embedded $w-$learning such that we present its
 analytical financial one- and two-rogon solutions with some free parameters. Moreover, we also illustrate
their dynamical behaviors for chosen different parameters $\alpha,\, \beta,\, k,\, a$, and $b$ (see Figs.~\ref{fig:1}-\ref{fig:4}).

 Our results may play an important role to explain some real financial crisis/storms (e.g., 1997 Asian financial crisis/storm and the current global financial crisis/storm).
Moreover, these results may further excite the possibility of relative researches and potential applications for
the financial rogue-wave phenomena in the financial markets and other related fields of science.

\vspace{0.15in}
 \noindent \textbf{Acknowledgements}

\vspace{0.01in}

 The work was partially supported by the NSFC60821002/F02 and NSFC11071242.



\end{document}